\documentclass[prb,floatfix,a4paper,twocolumn,superscriptaddress,showpacs,amsfonts,amssymb,amsmath]{revtex4-1}

\usepackage{notes2bib}
\usepackage{graphicx}
\usepackage{amsmath}
\usepackage{color}
\usepackage{enumitem}

\begin{document} 

\title{Energy splitting of image states induced by the surface potential corrugation of InAs(111)A}

\author{Jes\'us Mart\'inez-Blanco}
\affiliation{Paul--Drude--Institut f\"ur Festk\"orperelektronik, Hausvogteiplatz 5-7, 10117 Berlin, Germany}
\email{blanco@pdi-berlin.de}
\author{Steven C. Erwin}
\affiliation{Center for Computational Materials Science, Naval Research Laboratory, Washington DC 20375, USA}
\author{Kiyoshi Kanisawa}
\affiliation{NTT Basic Research Laboratories, NTT Corporation, Atsugi, Kanagawa, 243--0198, Japan}
\author{Stefan F\"olsch}
\affiliation{Paul--Drude--Institut f\"ur Festk\"orperelektronik, Hausvogteiplatz 5-7, 10117 Berlin, Germany}

\keywords{Scanning tunnelling microscopy, Scanning tunnelling spectroscopy, InAs, Field Emission Resonances}

\pacs{73.20.At, 74.55.+v, 68.37Ef, 68.47Fg}

\begin{abstract}
By means of scanning tunneling spectroscopy (STS) we study the electronic structure of the III-V semiconductor surface InAs(111)A in the field emission regime (above the vacuum level). At high sample bias voltages (approaching $+10$ V), a series of well defined resonances are identified as the typical Stark shifted image states that are commonly found on metallic surfaces in the form of field emission resonances (FER). At lower bias voltages, a more complex situation arises. Up to three double peaks are identified as the first three FERs that are split due to their interaction with the periodic surface potential. The high corrugation of this  potential is also quantified by means of density functional theory (DFT) calculations. Another sharp resonance not belonging to the FER series is associated with an unoccupied surface state.
\end{abstract}

\date{\today}
\maketitle

\section{Introduction}
At any solid--vacuum interface in which a change of polarizability takes place, a series of Rydberg--like unoccupied electronic states exists. They are called \textit{image states} and result from the coulombic tail of the potential in the vacuum side. Their energies lie below the vacuum level \cite{Echenique1978,Echenique1989,Echenique2002} and they are physically located in the near--surface region. Their wave functions are confined in the direction perpendicular to the surface, but disperse freely in the direction parallel to the surface. There are a number of suitable experimental techniques to detect them, like low--energy electron diffraction \cite{Echenique1978} or inverse photoemission \cite{Smith1988}. By means of scanning tunneling spectroscopy (STS) these states can also be mapped with atomic spatial resolution. The electric field created by the STM tip strongly modifies the interface potential, and the energy at which these states are detected is shifted to the field emission regime (above the vacuum level) due to the Stark effect \cite{Binnig1985}. 

The existence of these states has been exploited in the past for many purposes, like obtaining chemical contrast \cite{Jung1995}, measuring work function fluctuations \cite{Ruffieux2009}, achieving atomically resolved scanning tunneling microscopy (STM) images on insulators \cite{Bobrov2001} or even as qubits for quantum computation \cite{Platzman1999}. In recent years, increasing efforts have been made to understand how these electronic states are modified in the presence of surface nanostructures \cite{Schouteden2009,Schouteden2012,Stepanow2011,Polei2012}, steps and surface defects \cite{Pascual2007,Wahl2003} or rippled graphene, \cite{Borca2010,Zhang2010} to give just a few examples.

Although much experimental and theoretical research has been conducted on image states on metals, comparatively few studies have been performed on surfaces of semiconductors \cite{Kubby1990,Manghi1990,Polei2012,Xue2013}. In the present work, we employ STM and STS to study the rich electronic structure in the field emission regime of the III-V semiconductor surface InAs(111)A, and identify its image states from their observed field emission resonances (FER), which we analyze using a simple one--dimensional model potential defined between the surface and the STM tip. 

The InAs(111)A surface exhibits a $2\times 2$ indium-vacancy reconstruction which corrugates the ideal $1\times 1$ surface. Using density functional theory (DFT) we quantify the resulting corrugation of the electrostatic potential near the surface. A consequence of this corrugation is that the lower--order image states, whose electrons are confined closer to the surface, appear as split double resonances due to their interaction with the periodic potential of the surface. This feature can be understood in terms of scattering of the image state electrons by the corrugated surface potential, which prevents them from dispersing freely in contrast to their behavior at metal surfaces.

\section{Experimental and theoretical methods}
The experiments were carried out in an ultra--high vacuum chamber equipped with a low--temperature STM (\textit{Createc GmbH}) operated at 5 K. The substrate was prepared by growing 20 nm--thick undoped InAs layers on top of an InAs(111)A wafer (from \textit{Wafer Technology Ltd.}) by means of molecular beam epitaxy (MBE) monitored by \textit{in-situ} reflection high energy electron diffraction (RHEED). To enable sample transfer to the UHV system of the STM apparatus under ambient conditions, a capping layer of amorphous arsenic was deposited, and later desorbed by annealing at 630 K right before inserting the crystal into the STM head. We used a standard etched polycrystalline tungsten STM tip prepared in vacuum by front sputtering and annealing. Once the tip was in the STM stage, we performed repeated current pulses at sample bias voltages of up to 10 V. This treatment gives rise to an agglomeration of indium at the tip apex as evident from the tip behavior in subsequent atom manipulation experiments \cite{Yang2012}. The high voltage treatment was usually followed by gentle tip--surface contact to form a final atomically sharp tip apex. The differential conductance signal $dI/dV$ ($I$ denotes the tunneling current and $V$ the voltage) was measured with a lock--in amplifier using a bias oscillation frequency of 675 Hz and an amplitude of 40 mV. All bias voltages are referred to the sample with respect to the tip. 

First--principles DFT calculations were used to determine the equilibrium geometry and near--surface electrostatic potential of the InAs(111)A-$(2\times 2)$ surface. Total energies and forces were calculated within the generalized--gradient approximation of Perdew, Burke and Ernzerhof functional (PBE) \cite{Perdew1996} to DFT, using projector--augmented--wave (PAW) potentials as implemented in VASP \cite{Kresse1996}. The plane--wave cutoff for all calculations was 250 eV.  The surface calculations were performed in a slab geometry with ten layers of InAs and a vacuum region of 20 {\AA}. The topmost three atomic layers were relaxed until the largest force component on every atom was below 0.01 eV/{\AA}. The sampling of the 2$\times$2 surface Brillouin zone was carried out with a 3$\times$3 Monkhorst--Pack mesh centered at the $\Gamma$ point. 

\section{Results and Discussion}
\begin{figure}
\includegraphics[width=1\columnwidth]{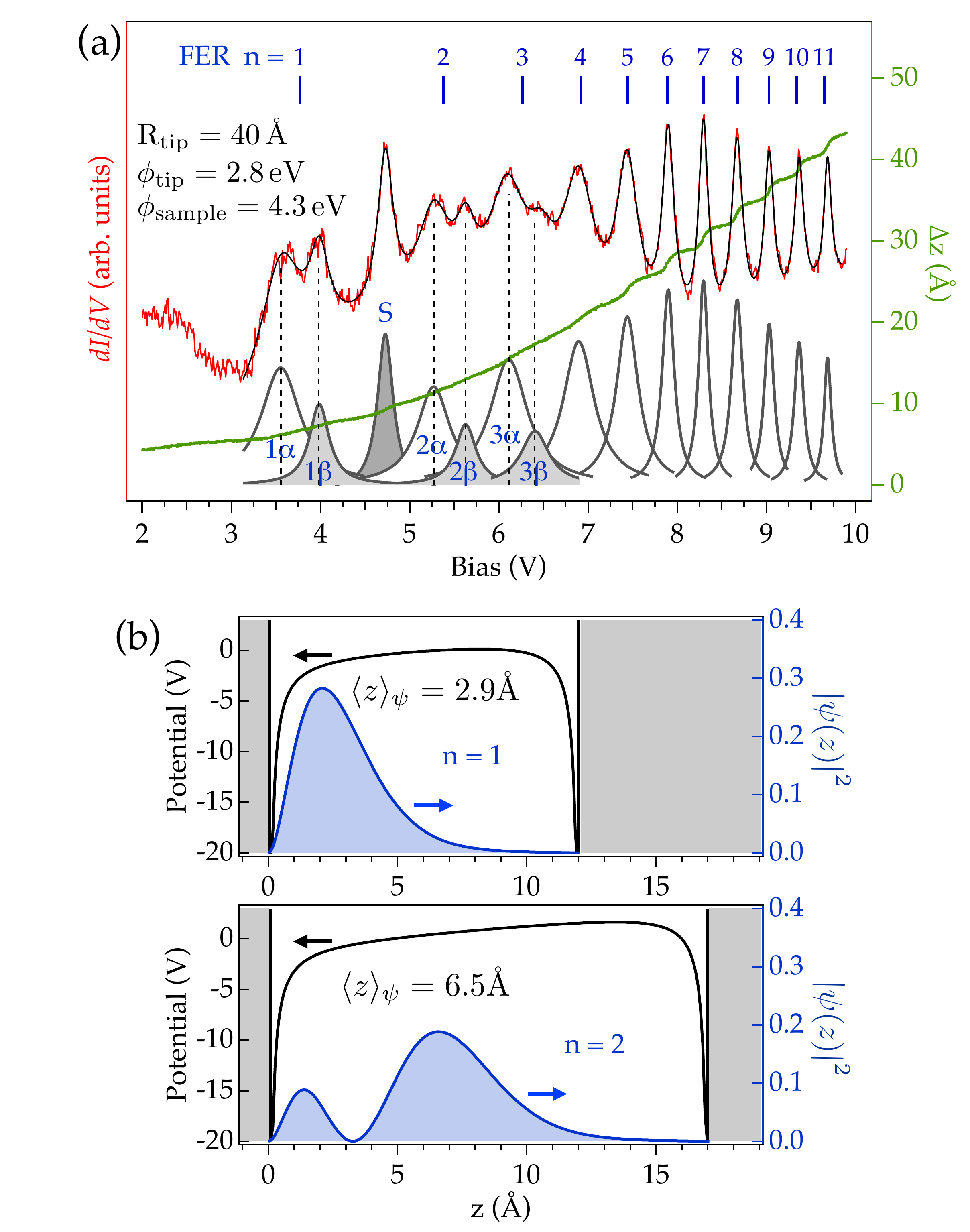}
\caption{(Color online) (a) Differential tunnel conductance $dI/dV$ (red curve) recorded on bare InAs(111)A as a function of the bias voltage and at a constant current of 8 pA. A fit of the spectrum to 15 lorentzian peaks plus a polynomial background is also shown (black curve). The tip retraction with respect to the initial tip--sample distance of about 5.5 {\AA} is shown in green. The blue vertical lines mark the location of the field emission resonances (FER) as calculated by solving the one--dimensional Schr\"odinger equation for the potential depicted in (b), using the indicated values for the parameters R$_\text{tip}$, $\phi_\text{tip}$ and $\phi_\text{sample}$. (b) One--dimensional potential model for two different tip heights, corresponding to the bias at which the tunnel junction is in resonance with the FERs $n=1$ and $n=2$. The associated electron probability densities are shown in blue. The corresponding expectation value for the electron position along the surface--tip axis $\langle z\rangle_\psi$ is also indicated.}
\label{fers}
\end{figure}

The red curve in Fig.\,\ref{fers}(a) corresponds to the differential conductance signal $dI/dV$ measured on InAs(111)A for bias voltages between $+2$ and $+10$ V. To prevent current overload as the bias voltage is ramped up into the field emission regime, the spectrum was acquired at constant current (feedback loop switched on), so that the tip height was changing continuously. The tip retraction was acquired simultaneously and is represented in green in this panel. From $+3$ to $+10$ V bias, 15 peaks are clearly visible in the spectrum, and our aim is to identify their nature. On top of the spectrum, a black curve marks the fit of the red curve to 15 lorentzian peaks with a cubic polynomial background. The individual fitted peaks are also shown in their corresponding locations.

A first inspection reveals that the identification of all peaks is not trivial, especially in the low bias range, where the energy separation between consecutive peaks does not seem to follow a clear pattern. To analyze this spectrum we develop here a simple model to help us in discerning whether a particular peak assignment is reasonable or not. The aim is to know where to expect the presence of the different FERs on InAs(111)A along an energy resolved spectrum such as the one shown in Fig.\,\ref{fers}(a). We solve the time--independent Schr\"odinger equation for an electron confined in the one--dimensional potential barrier shown as a black curve in Fig.\,\ref{fers}(b). The integration was done numerically\cite{Martinez-Blanco2015} using the Numerov method \cite{Numerov1927}. The potential is the sum of three contributions:
 
\begin{enumerate}[label=(\textit{\roman*})]
\item{the surface image potential,
\begin{equation} 
   \frac{-1}{4\pi\epsilon_0}\frac{e^2}{4}\left(\frac{\epsilon-1}{\epsilon+1}\right)\frac{1}{z},
   \label{SIP}
\end{equation}}
\item{the tip image potential,
\begin{equation} 
   \frac{-1}{4\pi\epsilon_0}\frac{e^2}{4}\left(\frac{1}{Z_0+Z(\text{V})-z}\right)
   \label{TIP},
\end{equation}
and}
\item the electrostatic potential between sample and tip. Assuming a spherical tip, the value of this potential along the cylindrical symmetry axis (coordinate $z$, representing the distance from the surface plane in the direction to the tip) can be calculated as described in Ref.[\onlinecite{DallAgnol2009}] and depends on the effective radius R$_\text{tip}$, the bias voltage, the tip--sample separation and the difference in work function between sample and tip, $\phi_\text{sample}-\phi_\text{tip}$.
\end{enumerate}

In expression (\ref{SIP}), $e$ is the electron charge, $\epsilon_0$ is the vacuum permitivity and $\epsilon$ is the sample permitivity, which has a value of 15.15 for InAs\bibnote{\url{http://www.ioffe.ru/SVA/NSM/Semicond/InAs/basic.html}}. In expression (\ref{TIP}), $Z(V)$ is the experimental tip displacement (green curve in Fig.\,\ref{fers}(a)) with respect to the initial tip--sample distance $Z_0$ corresponding to a set point of 0.1 V (initial bias in the spectrum) and 0.1 nA (constant current during $Z(V)$ spectroscopy). $Z_0$ has an estimated value\bibnote{The starting tip--sample distance is assumed to be the tip approach necessary to detect a current that deviates from the tunnelling exponential behaviour (not shown).} of 5.5 {\AA}. We choose the absolute zero of the potential to be located at the vacuum level of the sample. The potential is assumed to be infinite at both sides of the vacuum gap (hard--wall boundary conditions), for which we neglect the bulk potential in this approximation\cite{Ruffieux2009}. This assumption is equivalent to forcing the wave function to be zero at both sides of the tip--sample gap. Since image states are confined states in the $z$ direction mainly present in the tip--surface gap, we expect our model to provide a suitable scenario for fitting the experimental data.

This model has three free parameters, namely the sample and tip work functions and the radius of the tip. In the simulation, the value of these parameters was adjusted so that the maximum number of resonances can be explained. The vertical lines in Fig.\,\ref{fers}(a) show the fitted FER positions thus obtained using $\phi_\text{sample}=4.3$ eV, $\phi_\text{tip}=2.8$ eV and R$_\text{tip}=40$ {\AA}. The fit reproduces well the position of the last eight measured resonances and yields reasonable agreement for the first three, which we assume to be double resonances (labeled n$\alpha$ and n$\beta$, where n $=1,2,3$). The sharp resonance labeled S does not belong to the calculated series of FERs and is discussed later as well as the origin of the FER splittings. Although other peak assignments might be conceivable (for example, bulk related states), we discuss only the scenario in which these resonances are derived from image potential states, provided that a simple potential model yields a reasonable agreement with the data.

The experimental work function of InAs(111)A has not been reported in the literature. Work function values for III-V compound semiconductor surfaces typically vary between $4.4$ and $5.4$ eV (Refs.[\onlinecite{Fischer1967,Melitz2010}]). For example, the reported value for the As--terminated surface InAs(111)B is 4.7 eV (see Ref.[\onlinecite{Szamota-Leandersson2006}]). Therefore, the sample work function of $4.3$ eV obtained in the fit appears reasonable. On the other hand, the unrealistically small tip work funtion of $2.8$ eV and the tip radius of $\sim$40 {\AA} should be considered effective values arising from the simplifying approximations of the underlying model. 


To improve our understanding of the observed electronic structure, we modified the local electrostatic potential above the surface to see its effect on the different resonances. On InAs(111)A, this can be achieved by using charged defects such as indium adatoms (In), which are natural electron donors with a charge state $+1$ (Ref.[\onlinecite{Yang2012}]). In adatoms (either native or artificially created by soft tip indentation) can be readily repositioned by the STM tip\cite{Folsch2009,Yang2012}. Due to the low screening provided by the semiconductor surface, the In adatom charge is strongly localized and by repositioning and arranging them we can tailor the local electrostatic potential.

\begin{figure}
\includegraphics[width=1\columnwidth]{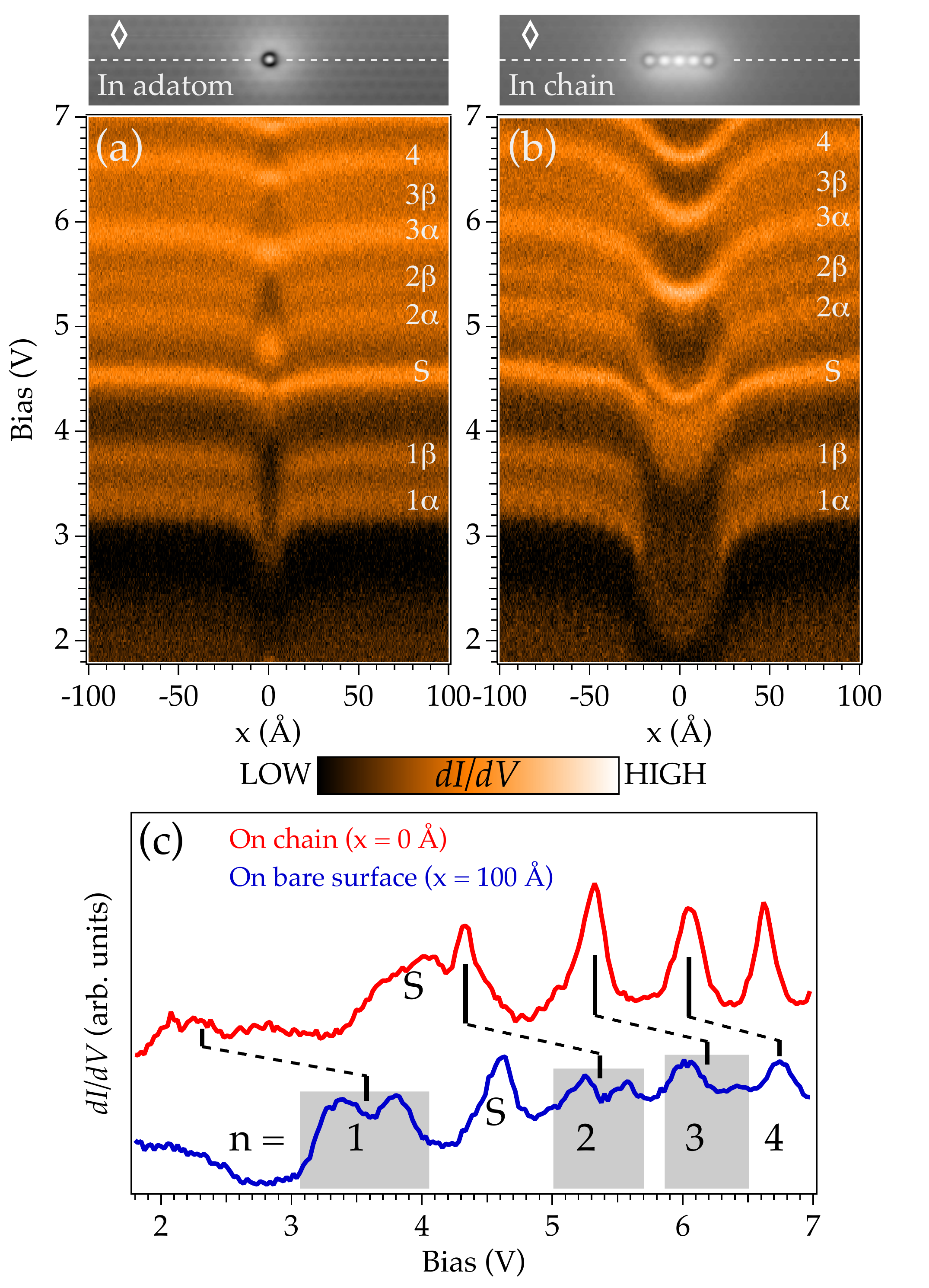}
\caption{(Color online) (a), (b) Differential conductance maps (constant current 10 pA) measured as a function of bias voltage and position along the white dashed line marked in the corresponding STM image shown above (0.5 V, 0.1 nA). The dashed line runs across a single In adatom (a) and a 5--membered adatom chain (b). The surface unit cell is marked with a white rhombus in the STM maps. c) Line profiles extracted from the map in (b) at $x=0$ {\AA} (in red) and at $x=100$ {\AA} (in blue). The curves are offset along the vertical axis for clarity.}
\label{nanos}
\end{figure}

The differential conductance maps in Fig.\,\ref{nanos}(a,b) show how the different resonances described for the bare surface in Fig.\,\ref{fers}, are affected in the presence of two nanostructures: a single adatom and a 5--membered adatom chain. The maps detail the bias dependence of $dI/dV$ profiles measured along the dashed lines in the STM images at the top of Fig.\,\ref{nanos}(a,b) respectively. Due to the larger charge associated with the chain, its effect is more dramatic. On the basis of this experiment, a number of observations can be stated. On the one hand, every double resonance measured on the bare surface (labeled with numbers 1, 2 and 3) is merged into a single enhanced resonance when measured on top of the nanostructure. The effect can be seen more clearly in Fig.\,\ref{nanos}(c) where the profiles extracted from Fig.\,\ref{nanos}(b) at $x=0$ {\AA} and at $x=100$ {\AA} are displayed for direct comparison. We observe also that all resonances are shifting to lower binding energies, as expected for the lowering of the local potential due to the positive charge provided by the nanostructure. This shift is also visible in the case of resonance S. However, contrary to the trend found for the FERs, the peak marked with S becomes broader and less intense on top of the nanostructure, confirming its distinct nature with respect to the rest of states in the series. The electronic structure of InAs(111)A in the unoccupied region has been studied with inverse photoemission spectroscopy (IPES) (see Ref.[\onlinecite{Ichikawa1998}]). According to that study, a surface state exists at $\sim 4.6$ eV with respect to the Fermi level of the surface. This value is fully compatible with the binding energy of resonance S obtained here with STS. We therefore conclude that S cannot be associated with an image state, and should be instead assigned to an intrinsic surface state, that broadens in the presence of the nanostructure.

\begin{figure}
\includegraphics[width=1\columnwidth]{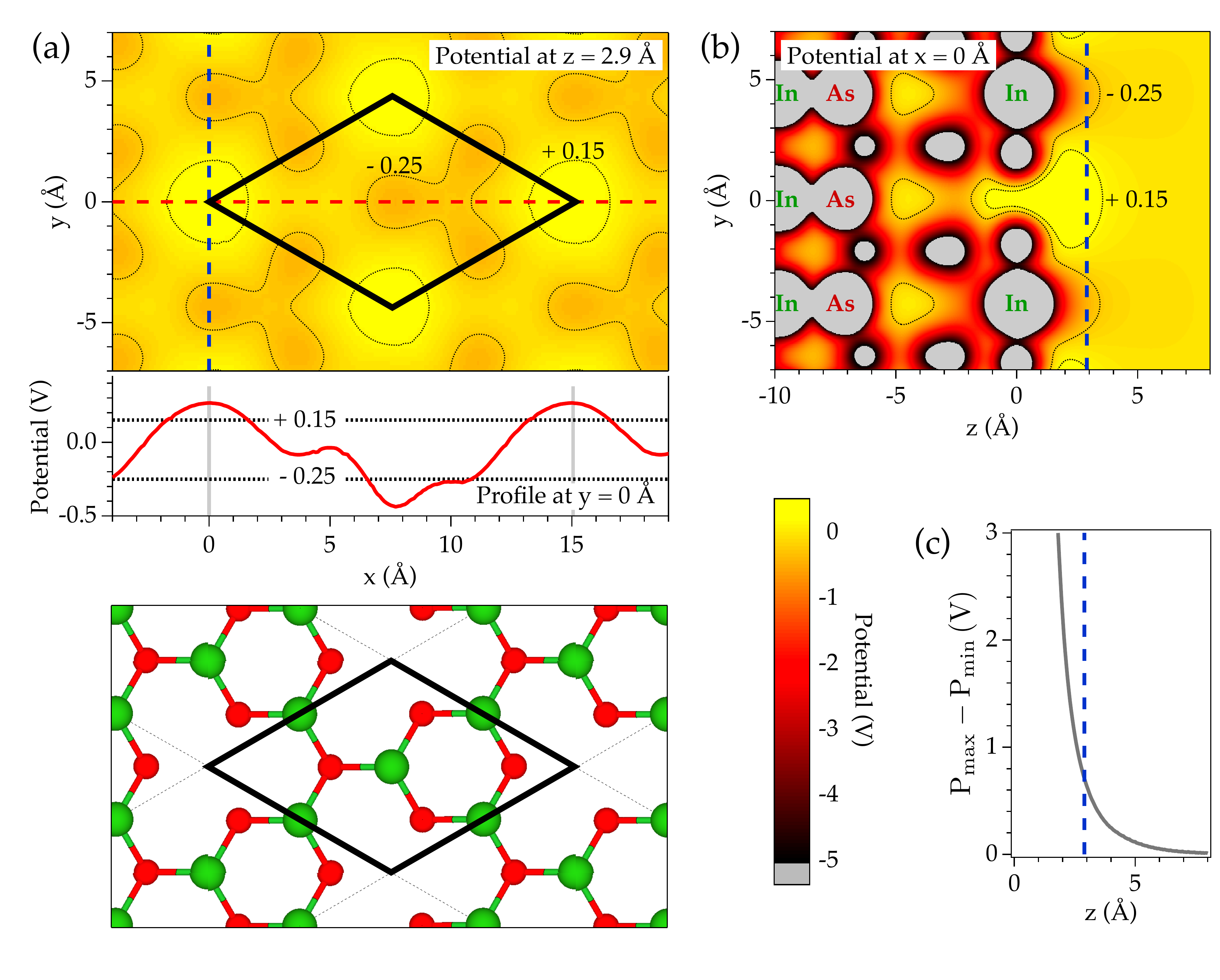}
\caption{(Color online) (a) DFT calculated potential map for InAs(111)A at a plane located $2.9$ {\AA} above the topmost atomic layer of the surface. The black rhombus marks the surface unit cell, and each of its vertices is located at the intrinsic vacancy sites that give rise to the surface reconstruction with $(2\times 2)$ periodicity. The red curve shown below is the profile across the center of the unit cell, as marked in the map with a red dashed line. The black dotted lines mark the isocontours for two values of the potential, $-0.25$ V and $+0.15$ V for a better visualization of the potential topology in the $xy$ plane. Below, the stick--and--ball model of the topmost bilayer of atoms (indium depicted in green and arsenic in red) helps to identify the surface orientation and locations. (b) DFT calculated potential map within the $x=0$ plane, which intersects the map in (a) along the dashed blue line. The intrinsic vacancy is located at $(z,y)=(0,0)$, where an In atom is missing in the last atomic layer. The isocontours for $-0.25$ V and $+0.15$ V are also represented. (c) Potential corrugation ($\text{P}_\text{max}-\text{P}_\text{min}$) for every plane $z$, on the vacuum side. The potential is still considerably corrugated for relatively large values of $z$. The blue dashed line in (b) and (c) corresponds to the plane $z=2.9$ {\AA} depicted in (a).}
\label{surfpot}
\end{figure}

Next, we address the origin of the splitting observed for the lower--order FERs. The associated electron probability densities for the first two FERs is shown in blue in Fig.\,\ref{fers}. These curves represent the calculated electron distribution along the $z$ direction (perpendicular to the surface) in the sample--tip gap. Note that in the 1D model used to obtain these distributions, the FER electrons disperse freely in the direction parallel to the surface but are confined in the $z$ direction. The expectation value for the electron position along the surface--tip axis $\langle z\rangle_\psi$, is $2.9$ {\AA} for the FER with $n=1$. This value is quite close to the surface, raising the question of whether the crystal potential is sufficiently smooth at $z=2.9$ {\AA} to ensure free dispersion of the FER wavefunction parallel to the surface. In fact, we find that the potential is quite corrugated. Figure \ref{surfpot}(a) shows the full DFT electrostatic potential $2.9$ {\AA} above the surface. Fig.\,\ref{surfpot}(b) shows a cross--section of the potential containing the surface normal and the in--plane direction at $x=0$ marked by the vertical dashed line in Fig.\,\ref{surfpot}(a). The In and As atoms in this plane are labeled. The intrinsic vacancy is located at $(z,y)=(0,0)$, where an In atom is missing in the topmost atomic layer. The red curve shown below the potential map in Fig.\,\ref{surfpot}(a) shows the potential along the horizontal dashed line. This profile runs across the positions of the maximum and minimum in the entire potential map, corresponding to the intrinsic vacancy sites and the In surface atoms respectively. A stick--and--ball model of the surface bilayer is shown below.

The corrugation of the DFT potential at any given $z$ plane, which we define as the difference between its maximum and minimum ($\text{P}_\text{max}-\text{P}_\text{min}$), is very large near the surface ($0.7$ V at $z=2.9$ {\AA}) and rapidly decreases at larger $z$, as shown in Fig.\,\ref{surfpot}(c). The electronic states belonging to the first FERs have a significant presence in the near--surface region, where the potential corrugation is still high, and thus undergo scattering by the periodic potential in the direction parallel to the surface. Due to this interaction, the low--order states are energetically split. If the interaction is strong enough, the energy dispersion of the image state  in the plane parallel to the surface can deviate significantly from the free--electron like dispersion assumed in our 1D model\cite{Manghi1990}.


\begin{figure}
\includegraphics[width=1\columnwidth]{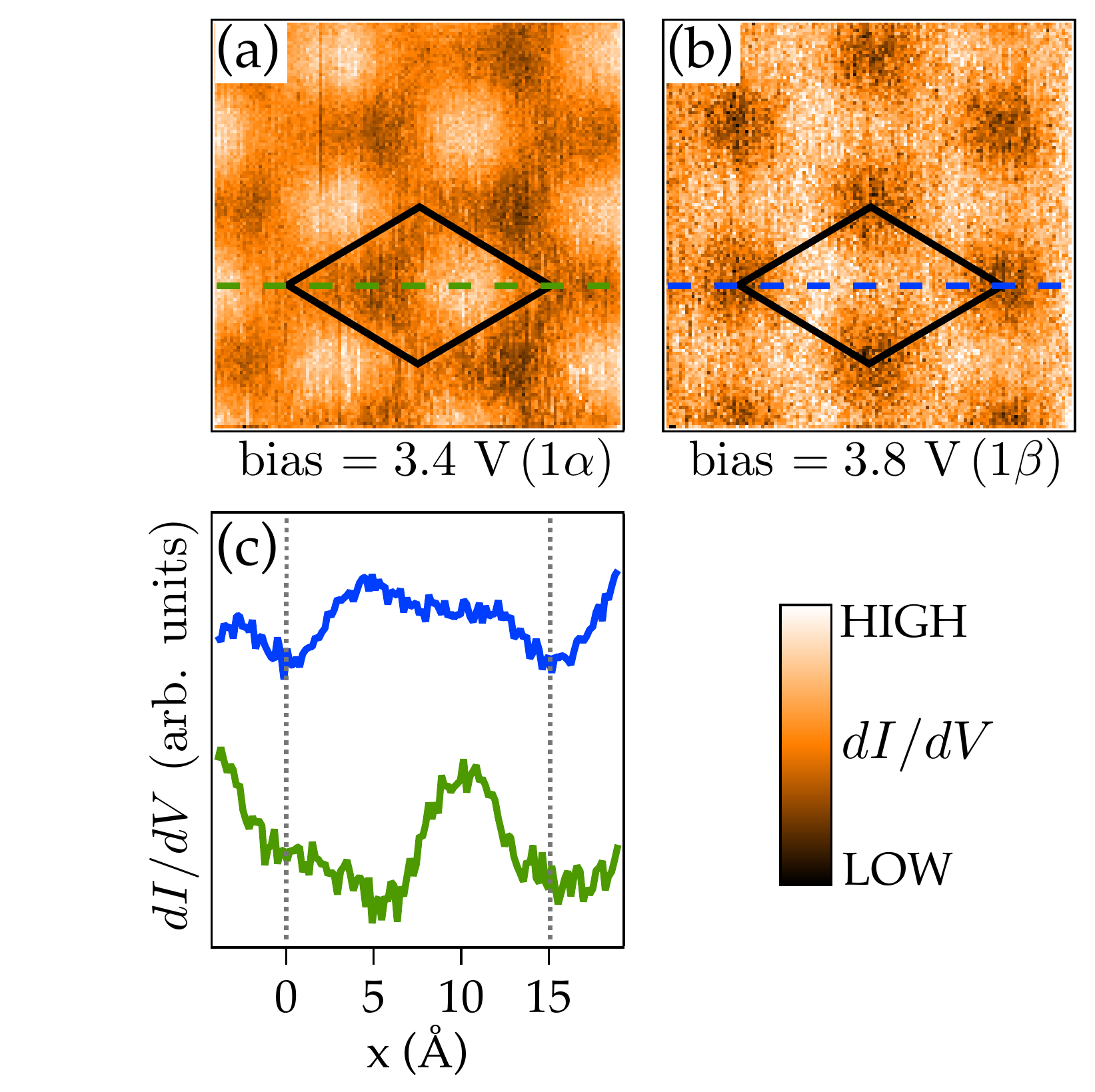}
\caption{(Color online) (a), (b) Constant height differential conductance maps on InAs(111)A for the bias voltages corresponding to resonances 1$\alpha$ and 1$\beta$ shown in Fig.\,\ref{fers}(a), respectively. The surface unit cell is marked with a black rhombus. (c) Line profiles along the dashed lines indicated in panels (a) (in green) and (b) (in blue) showing different LDOS distributions across the surface unit cell.}
\label{maps}
\end{figure}

This reasoning is supported quantitatively by the data in Fig.\,\ref{maps}. Panels (a) and (b) show spatial maps of the differential conductance measured on the bare surface for two different bias voltages, corresponding to the FERs $1\alpha$ and $1\beta$, respectively. Each of the four vertices of the black rhombus marking the surface unit cell, is located at an intrinsic vacancy site, and both Fig.\,\ref{surfpot} and Fig.\,\ref{maps} display the same surface orientation\bibnote{The 2$\times$2 reconstruction of InAs(111)A has three possible orientations and the surface is divided in domains of each of these orientations. Due to the broken bonds of the surface In and As atoms along a domain border, STM imaging at both sides of the border is distinct in its vicinity. We use this topographic fingerprint to identify the surface orientation.}. Because the maps of Fig.\,\ref{maps}(a) and (b) were measured at constant tip height, they provide a measure of the lateral distribution of electronic states at the particular energies associated with each of the two peaks of the first FER doublet. The lower--energy peak $1\alpha$ has most of its electronic density concentrated at the center of the right half of the unit cell (Fig.\,\ref{maps}(a)). This region coincides with the surface location at which the potential deepens, as can be seen in Fig.\,\ref{surfpot}(a). The isocontour labeled with $-0.25$ encloses the region in which the potential is lower than $-0.25$ V and where the absolute minimum is located. In contrast, the state density associated with the higher energy peak $1\beta$ (Fig.\,\ref{maps}(b)) distributes mainly at the center of the left half of the unit cell, but extends over the rest of the surface with lower density at the vacancy sites, forming a honeycomb structure. In Fig.\,\ref{surfpot}(a), the region delimited by the isocontours at $-0.25$ and $+0.15$ V extends over the same region of the $xy$ plane as the honeycomb structure. Hence, the probability density of the lower (higher) energy state is concentrated in the regions where the potential is lower (higher). This situation resembles the simple case of gap formation in the nearly--free--electron model, where eigenstates at the Brillouin zone boundary have a high expectation value either between (higher energy) or centered around the nuclei positions (lower energy) defining the minima of the periodic potential. A completely free--electron image state would instead have a constant probability density across the $xy$ plane, as commonly found on metal surfaces.

We described in Fig.\,\ref{nanos} how the double resonances observed on the bare surface become single peaks when measured on the adatom nanostructures. In the context of our model, this phenomenon can be understood as an indication that the potential corrugation is smoother on the nanostructures as compared to the bare surface.

\section{Conclusions}
We have studied the electronic structure of the InAs(111)A surface in the field emission regime and gained a qualitative, as well as semi--quantitative understanding of the resonances observed in scanning tunneling spectroscopy. In contrast to the behavior observed on metal surfaces, our results reveal a sizable splitting of the lower--order field emission resonances (up to $n=3$) in the low--bias regime. We interpret this splitting as an indication that the lower--order surface image states do not disperse freely in the direction parallel to the surface, due to the relatively large surface potential corrugation predicted by DFT calculations. The spatial distribution of the electronic states observed at the respective energies of the split resonances confirms this picture. In addition to the field--emission--resonance states, the spectroscopic data reveal an electronic state that does not belong to the series of image states and is assigned to an intrinsic surface state of InAs(111)A. Our results show that the surface potential corrugation inherent to a compound semiconductor surface can have significant impact on the energy level spectrum of image states in the field--emission regime.   

\section{Acknowledgement}
This work was supported by the German Research Foundation (SFB 658, TP A2) and the Office of Naval
Research through the Naval Research Laboratory's Basic Research Program. 
Computations were performed at the DoD Major Shared Resource Center at 
AFRL.


%

\end{document}